\documentclass[prb,showpacs,twocolumn,superscriptaddress]{revtex4}
\usepackage{epsfig}
\usepackage{amsmath}
\usepackage{graphicx}   
\usepackage{dcolumn}    
\usepackage{longtable}
\usepackage{bm}         
\usepackage{color}
\usepackage{multirow}

\definecolor{darkgreen}{rgb}{0., 0.6, 0.2}

\usepackage{gensymb}

\begin{document}

\draft
\title{Ab-initio based models for temperature-dependent magneto-chemical interplay in bcc Fe-Mn alloys}

\author{Anton Schneider}
\affiliation{Universit\'e Paris-Saclay, CEA, Service de Recherches de M\'etallurgie Physique, 91191 Gif-sur-Yvette, France}

\author{Chu-Chun Fu}
\affiliation{Universit\'e Paris-Saclay, CEA, Service de Recherches de M\'etallurgie Physique, 91191 Gif-sur-Yvette, France}

\author{Osamu Waseda}
\affiliation{Max-Planck-Institut f\"ur Eisenforschung, D\"usseldorf, Germany}

\author{Cyrille Barreteau}
\affiliation{DRF-Service de Physique de l'Etat Condens\'e, CEA-CNRS, Universit\'e Paris-Saclay, F-91191 Gif-sur-Yvette, France}
 
\author{Tilmann Hickel}
\affiliation{Max-Planck-Institut f\"ur Eisenforschung, D\"usseldorf, Germany}

\date{\today}

\begin{abstract}
Body-centered cubic (bcc) Fe-Mn systems are known to exhibit a complex and atypical magnetic behaviour from both experiments and 0 K electronic-structure calculations, which is due to the half-filled 3d-band of Mn. 
We propose effective interaction models for these alloys, which contain both atomic spin and chemical variables. They were parameterized on a set of key density functional theory (DFT) data, with the inclusion of non-collinear magnetic configurations being indispensable. Two distinct approaches, namely a knowledge-driven and a machine-learning approach have been employed for the fitting.
Employing these models in atomic Monte Carlo simulations enables the prediction of magnetic and thermodynamic properties of the Fe-Mn alloys, and their coupling, as functions of temperature.  
This includes the decrease of Curie temperature with increasing Mn concentration, the temperature evolution of the mixing enthalpy and its correlation with the alloy magnetization. Also, going beyond the defect-free systems, we determined the binding free energy between a vacancy and a Mn atom, which is a key parameter controlling the atomic transport in Fe-Mn alloys.

\end{abstract}

\maketitle

\section{Introduction}
\label{intro}
Thermodynamic and kinetic properties of Fe-based alloys can be strongly influenced by magnetism. For instance, previous studies have shown that magnetism in Fe-Cr alloys has a crucial impact on the mixing enthalpy and induces an asymmetry in the mutual solubility of Fe and Cr at low temperature \cite{Ruban2012, Senninger2014}. It is also believed that, for a large range of concentrations in Fe-Co alloys, the ferromagnetic-paramagnetic (FM-PM) transition is closely linked to the body-centered-cubic (bcc) to face-centered-cubic (fcc) structural transition \citep{Nishizawa1990,Ohnuma2002,Sourmail2005}. In fcc Fe-Mn alloys stacking-fault energies strongly depend on magnetic order \citep{Dick2009}. Also, it is clearly known from experiments that the FM-to-PM transition in a bcc Fe system induces an abrupt acceleration of the diffusion of Fe and most of solute \citep{Luebbehusen1984,Kucera1974,Graham1963,Geise1987,Borg1960,Walter1969,Iijima1988,Hettich1977,Buffington1961,Huang2010, Schneider2020}. Among the binary Fe alloys, the Fe-Mn system exhibits some special behavior related to the half-filled $3d$-band of Mn. For example, the typical diffusion acceleration around the Curie point is not visible for Mn solutes in bcc Fe \citep{Kirkaldy1973,Luebbehusen1984}. Pure bulk Mn itself shows a complex magneto-structural phase diagram \cite{Hobbs2003,Hafner2003}. Further, a strong magneto-elastic coupling in Fe-Mn alloys was raised previously\cite{Music2007}. However, so far, the magneto-thermodynamic and magneto-kinetic interplays in Fe-Mn systems remain unclear.

This study is focused on bcc Fe-Mn, since we are mainly interested in the Fe-rich part of this alloy. As can be seen in the equilibrium phase diagram from, for example,  Witusiewicz {\it et al.} \citep{Witusiewicz2004} and Bigdeli {\it et al.} \citep{Bigdeli2019}, the stable domain of bcc Fe-Mn is limited to the dilute region (at most 5 at. \% Mn). Despite the restricted homogeneity region, bcc Fe-Mn presents some intriguing properties, such as an extremely environment-dependent magnetic state of Mn solutes \citep{Anisimov1988,Schneider2018} and the anomalous Mn-solute diffusion behavior in bcc Fe-Mn. An accurate finite-temperature modelling of  properties in this system that properly account for magnetic effects is highly necessary but challenging.

Up to now, mainly two distinct modelling approaches have been employed for the study of magnetic alloys at finite temperatures. On one hand, the disordered local moment (DLM) approach \citep{Alling2010a,Alling2010b,Ruban2007,Khmelevska2007,Dutta2012,Lagarec2001}, often combined with the Coherent Potential Approximation (CPA), allows one to describe alloys with both chemically and magnetically random structures, in order to mimic an ideal paramagnetic solid solution occurring at high temperatures. Recently, the itinerant coherent potential approximation (ICPA) has also been employed for the description of the PM state \citep{Dutta2020}. Some DLM studies were also carried out adopting supercells. For instance, a spin-space average approach was used to describe the disordered magnetic state \citep{Koermann2012, Ruban2012, Ikeda2014}. Note that magnetic short range orders (SRO) are usually not taken into account in these approaches, and that the temperature evolution of the alloy properties cannot be directly predicted.

The other methodology, based on the parameterization of effective interaction models (EIMs) containing a magnetic and a chemical contributions has also shown interesting results for the study of materials properties as functions of temperature. Namely, Pierron-Bohnes {\it et al.} have used such an approach, based on the Ising model, in an early study on Fe-Co alloys \citep{PierronBohnes1983}. A more sophisticated magnetic cluster expansion (MCE) approach was later applied by Lavrentiev {\it et al.} to the study of Fe-Ni \citep{Lavrentiev2014}, Fe-Cr \citep{Lavrentiev2010} and Fe-Ni-Cr \citep{Lavrentiev2016} alloys. Similar models were also developed by Ruban {\it et al.} \citep{Ruban2007} for Fe-Cr and by Tran {\it et al.} \citep{Tran2020} for Fe-Co alloys. These models are generally used in on-lattice Monte Carlo (MC) simulations for finite temperature studies,  Chapman {\it et al.} \citep{Chapman2019} have recently adapted the MCE model by Lavrentiev {\it et al.} \citep{Lavrentiev2010} for spin-lattice dynamics simulations. The latter simulation technique is exempted of the rigid-lattice constraint, but is  computationally much more demanding.

So far, such effective models have mostly been employed for the prediction  of magnetic properties rather than thermodynamic properties such as phase diagrams. The simulations for the former are usually performed at fixed atomic structures, where only the magnetic structure evolves. This is mainly due to the complexity and high computational cost of MC simulations if dealing with a coupled evolution of both chemical and magnetic degrees of freedom. Moreover, the available models are limited to defect-free alloys, which prevents from modeling kinetic properties.

In order to investigate the full  interplay between magnetism and thermodynamic and defects properties for bcc Fe-Mn alloys, we aim at developing an effective interaction model, fitted on ab-initio results, which takes  all the relevant chemical and magnetic variables explicitly into account. 
Because the Mn magnetic moments exhibit significant magnitude variations for different local chemical environments, and in a highly complex way, the magnetic-interaction part of our model is based on a generalized Heisenberg formalism \citep{Lavrentiev2010,Ruban2012}. 
We first obtained a model for ideal Fe-Mn alloys. Then, as real materials are never defect-free, we also modified the obtained defect-free EIM to include the presence of a vacancy, as previously done for bcc Fe \citep{Schneider2020}. Being the simplest of the structural defects, vacancies play a crucial role in the atomic transport in  Fe alloys. Note that magnetic properties of Fe and Mn atoms and the chemical interactions can significantly change around a vacancy \citep{Schneider2018}. 
 
 An accurate parameterization procedure of such EIMs is generally a non-trivial task. In this paper, we propose two different strategies: the first one relies on the knowledge of key properties of the system, identified from prior density functional theory (DFT) results. The second one applies a machine-learning approach. Both resulting models (respectively knowledge-driven (KD) and machine-learning (ML) models) are compared for quality assessment, and advantages and drawbacks of the two approaches are derived and discussed.   

The paper is organized as follows: Sec. \ref{comp_details} describes details of DFT calculations for the models parameterizations, and of Monte Carlo simulations for finite-temperature studies. The two strategies for the parameterization of EIMs are explained in Sec. \ref{eim}, and the accuracy of the obtained models are verified in Sec.  \ref{validation}. Then, the EIMs are applied (in Sec. \ref{results}) to the prediction of various temperature-dependent properties that cannot be accessed directly from DFT calculations. As much as possible, the agreement with experimental or Calphad results is discussed.

\smallbreak
\section{Computational methods}
\label{comp_details}

Throughout the paper, magnetic moments are expressed in Bohr magnetons and the model parameters are energies, expressed in meV.

\smallbreak
\subsection{Density functional theory calculations}
\label{dft_details}

In this work, density functional theory (DFT) calculations are performed in order to parameterize the effective interaction models. Although a full description of these calculations and the results is given in Ref. \onlinecite{Schneider2018}, we provide the key features in this section.

The DFT calculations are performed using the Projector Augmented Wave (PAW) method \cite{Bloechl1994,Kresse1999} as implemented in the VASP (Vienna Ab-initio Simulation Package) code\cite{Kresse1993,Kresse1996a,Kresse1996b}. The results presented are obtained using the generalized gradient approximation (GGA) for the exchange-correlation functional in the Perdew-Burke-Ernzerhof (PBE) form \cite{Perdew1999}. All the calculations are spin-polarized. $3d$ and $4s$ electrons are considered as valence electrons. The plane-wave basis cutoff is set to 400 eV. Atomic magnetic moments are obtained by a charge and spin projection onto the PAW spheres \citep{Kresse1996a,Kresse1996b} as defined by the PAW potentials.

The $k-$point grids used in our calculations were adjusted according to the size of the supercell. They were chosen to achieve a $k$-sampling equivalent to a bcc cubic unit cell with a $16\times16\times16$ shifted grid, following the Monkhorst-Pack scheme.\cite{Monkhorst1976} The Methfessel-Paxton broadening scheme with a 0.1 eV width was used.\cite{Methfessel1989} The convergence threshold for the electronic self-consistency loop was set to $\Delta$E = $10^{-6}$ eV and atomic relaxations at constant volume were performed down to a maximum residual force of 0.02 eV/\AA. We have verified that the magnetic structures and cluster formation energies are well converged with respect to the choice of $k-$point grids and the cutoff conditions.
The resulting error bars for energy differences and magnetic moments of Fe and Mn are respectively 0.02 eV, 0.01 $\mu_{\text{B}}$ and 0.1 $\mu_{\text{B}}$. These are mainly associated to the convergence of the plane-wave energy cutoff and the $k-$grid density.

All the alloy concentrations given in the paper are expressed as atomic percent, if not explicitly otherwise indicated.

Our fitting database consists in several systems containing Mn, in the form of isolated Mn solutes in the Fe lattice, Mn dimers at various distances, clusters from 2 to 15 Mn atoms and various random solid solutions over the whole range of concentration. For each of these systems, several magnetic configurations have been generated. We note that, for isolated Mn solutes and Mn dimers in pure Fe, various non-collinear magnetic configurations were also considered. Considering the various chemical and magnetic configurations, the fitting database contains approximately 20 dimer configurations, 20 Fe-Mn SQS systems and one hundred Mn-cluster configurations. For more details, see Ref. \onlinecite{Schneider2018}.

\smallbreak
\subsection{Monte Carlo simulations}
\label{mc_details}

All the Monte Carlo (MC) simulations presented in this work are performed on a 16000 atoms system assuming a bcc lattice (20x20x20 cubic unit cells). Random solutions are generated by randomly distributing $p$ Mn atoms in the Fe matrix, where $p$ is defined as the total number of atoms multiplied by the imposed Mn atomic concentration. For each simulation, $5\cdot 10^{8}$ initial spin equilibration Metropolis attempts are performed in order to thermalize the magnetic structure. In the most difficult case, the convergence is reached after approximately $10^{8}$ steps. For the spin-MC simulations at a fixed atomic configuration, We perform additional $4\cdot 10^{8}$ MC-steps for data collection.

In the simulations involving both magnetic equilibrations and atomic exchanges (namely spin-atom MC), after the spin thermalization, we perform a Metropolis  attempt to exchange the respective positions of two randomly chosen atoms, once every $N_s$ spin-MC steps.
Convergence tests were done in order to ensure that enough spin-MC steps are performed between two successive atom-exchanges attempts. In the present calculations, 100 spin-MC attempts are carried out randomly anywhere in the system and 500 spin-MC attempts are performed in the two nearest-neighbor shells of the exchanged atoms.

In the presence of a vacancy, we follow a Monte Carlo method proposed in our previous study \citep{Schneider2020} which allows to determine the vacancy formation magnetic free energy as a function of temperature. The overall principle is that two separate subsystems are considered with the first one frozen at the FM state, while the magnetic configuration of the second one is allowed to evolve according to temperature. The vacancy is allowed to visit each site of the two subsystems via the Metropolis algorithm. Note that when the vacancy goes from one subsystem to the other,  the resulting energy change should also account for the fact that the energy of an Fe atom, at each temperature,  is generally different in both subsystems.
Then, based on the relative number of “visits” to the two systems and the vacancy formation energy at the FM state, which is known to be 2.20 eV. More details can be found in Ref. \onlinecite{Schneider2020}.

Please note that the lattice vibrational entropies are not intrinsically accounted in the present EIM–Monte Carlo setup. When necessary, they can be calculated additionally using DFT, in a similar way as in Ref. \onlinecite{Schneider2020}.

\smallbreak
\section{Effective interaction model}
\label{eim}

In this study, the EIMs adopt a Hamiltonian within the same formalism as in Refs.~[\onlinecite{Lavrentiev2010,Lavrentiev2014}]. It is composed of a magnetic-interaction part, which includes a Landau-expansion term and a Heisenberg-like term, and a pairwise chemical (nonmagnetic) interaction part. 

The Hamiltonian has the following formal expression:
\begin{equation}
\begin{split}
\label{total_hamil}
H = \sum\limits_{i}^{N} (A_{i}M_{i}^{2} + B_{i}M_{i}^{4} + C_{i}M_{i}^{6}) \\ + \sum\limits_{i}^{N}\sum\limits_{n}^{P}\sum\limits_{j}^{Z_n} J_{ij}^{(n)}{\bf M}_{i}\cdot {\bf M}_{j} \\ + \sum\limits_{i}^{N}\sum\limits_{n}^{P}\sum\limits_{j}^{Z_n} V_{ij}^{(n)}
\end{split}
\end{equation}
where $P$ is the maximum range of interactions in terms of neighbor shells, $Z_n$ is the coordination number of each neighboring shell, ${\bf M}_{i}$ is the magnetic moment of the $i$-th atom, $M_{i}$ is its magnitude. $V_{ij}^{(n)}$ and $J_{ij}^{(n)}$ represent respectively the chemical pair-interaction and the magnetic exchange-coupling parameters  between atoms $i$ and $j$, at a range $n$. 

$A_{i}$, $B_{i}$ and $C_{i}$ are the magnetic on-site parameters of the $i$-th atom.
In order to keep this model as simple as possible, only three terms are considered in the Landau expansion.
Their role is to prevent the divergence of the magnetic moment magnitudes caused by the spin longitudinal variations due to the Heisenberg-like terms.

We use the DFT data described in Sec.~\ref{dft_details} to determine the free parameters. We present, in the following subsections, the two models fitted on the same DFT dataset and based essentially on the same Hamiltonian but resulting from the two different parameterization strategies:

\subsection{Knowledge-driven (KD) model}
\label{model I}

For this model, we have chosen to determine the free parameters guided by key characteristics of this system that are revealed by the DFT investigations. In particular (i) in the presence of a magnetic frustration, the Mn-Mn interaction generally dominates over the Fe-Mn  AF tendency, and (ii) the presence of two magnetic minima for a Mn solute, with their relative stability highly dependent on the local chemical environment\cite{Schneider2018}.  
 
 A least-squares fitting  method is applied. As the problem is over-determined, we applied a progressive parameterization procedure, to reproduce a large number of physical properties derived from the DFT calculations. To this end, a priority is given to those data we consider to be the most important and compromises are accepted for less relevant properties as, e.g., NM Fe.

The procedure consists in fitting first the magnetic-interaction parameters of pure Fe and pure Mn on the respective bulk magnetic properties. In a second step, we fit the Fe-Mn magnetic interactions parameters, keeping  the pure-system parameters fixed.
In order to fit these magnetic parameters, we use the energy difference between DFT systems with identical atomic configurations but distinct magnetic states. It allows to attribute the total energy difference to the variation of the magnetic state. As the EIM will be applied to on-lattice MC simulations, any deviation from the perfect bcc structure  due to atomic relaxations in the DFT calculations is not explicitly considered, but its effect on the total energy is taken into account in the energy differences.  

Finally, once a satisfactory set of magnetic parameters is found, the non-magnetic parameters (associated to chemical bonding) are fitted on the DFT predicted mixing energies of Fe-Mn random solutions, represented by special quasi-random structures (SQSs)

\subsubsection{Magnetic parameters from pure bcc Fe and bcc Mn properties}
\label{model_Fe}

The Fe-Fe magnetic exchange-coupling parameters ($J_{ij}$s) and the on-site parameters ($A_{\rm Fe}$ and $B_{\rm Fe}$, for the sake of simplicity $C_{\rm Fe}=0$) are determined by using energies from DFT calculations performed on pure Fe systems (128-atom supercells) with various magnetic states. The $J_{ij}$s are fitted up to the fifth nearest-neighbor ($5nn$) shell.
The DFT systems include ferromagnetic (FM), anti-ferromagnetic (AF), double-layer anti-ferromagnetic (AFD), quadruple-layer anti-ferromagnetic (AF4), non-magnetic (NM) and also tens of magnetically disordered systems (random collinear magnetic moments).
For the magnetically ordered structures, the calculations are performed using the corresponding equilibrium lattice constant ($a_{0}$). For the magnetically disordered structures, the FM equilibrium $a_{0}$ was assumed. We checked that the residual pressure remains lower than 10 kbar.
We note that the obtained set of $J_{ij}$s is highly consistent with an earlier literature study \citep{Pajda2001}.

Since we are mainly interested in the Fe-rich region of bcc Fe-Mn alloys, an accurate description of bcc Mn bulk properties is not a priority. However, it is still necessary to fit correctly the Mn magnetic parameters ($A_{\rm Mn}$, $B_{\rm Mn}$ and $J_{ij}$s, for the sake of simplicity $C_{\rm Mn}=0$) in order to predict properly the interaction between various Mn solutes and the Mn clustering in the Fe lattice. We fit these interaction parameters on DFT data of pure bcc Mn in a similar way as for the Fe parameters.
The $J_{ij}$s are also considered up to the fifth neighbor shell. 

Please note that the relative values of Fe-Mn and Mn-Mn $J_{ij}$s control the competition of these interactions in the presence of a magnetic frustration. They have a critical effect in the determination of the magnetic ground-state of Fe-Mn systems, especially when Mn clusters are present. Therefore, in practice, the obtained Mn-Mn parameters are slightly adjusted once the Fe-Mn $J_{ij}$s are determined.  

\begin{figure}[htbp]
  \centerline{\includegraphics[width=1.0\linewidth]{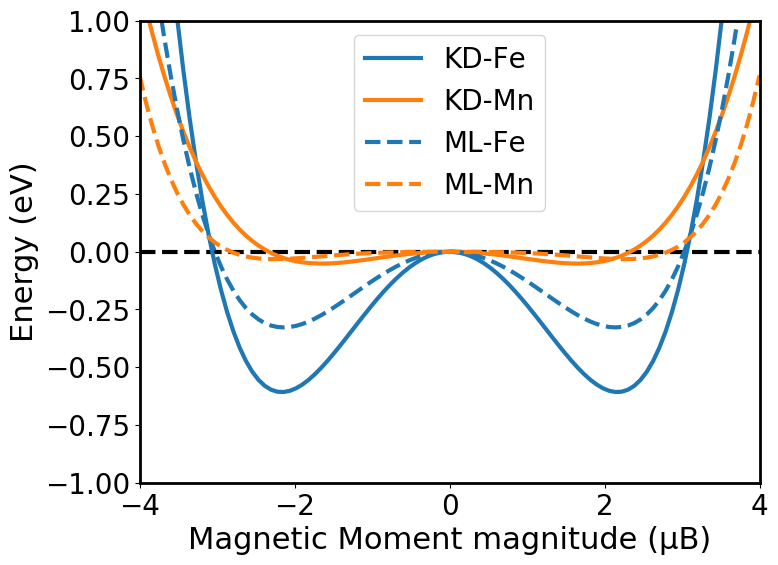}}
  \caption{Evolution of the magnetic on-site energy as a function of the magnetic moment magnitude. The magnitude of Fe spins is stiffer than the magnitude of Mn spins.}
  \label{abcomp}
\end{figure}

Fig. \ref{abcomp} shows the curves of the magnetic on-site energy imposed by the Landau expansion as a function of the magnetic moment magnitude for Fe and Mn atoms. The onsite energies $A_i$, $B_i$ and $C_i$ in Eq.~(\ref{total_hamil}) do not depend on the magnetic environment.
As can be noticed in Fig. \ref{abcomp}, the minimum of the Mn curve is more shallow and flatten  than the one of Fe. This is consistent with DFT data\cite{Schneider2018} indicating that the magnetic moment magnitude of Mn atoms is much more dispersed than the ones of Fe atoms, for different local environments.

\begin{figure}[htbp]
  \centerline{\includegraphics[width=1.0\linewidth]{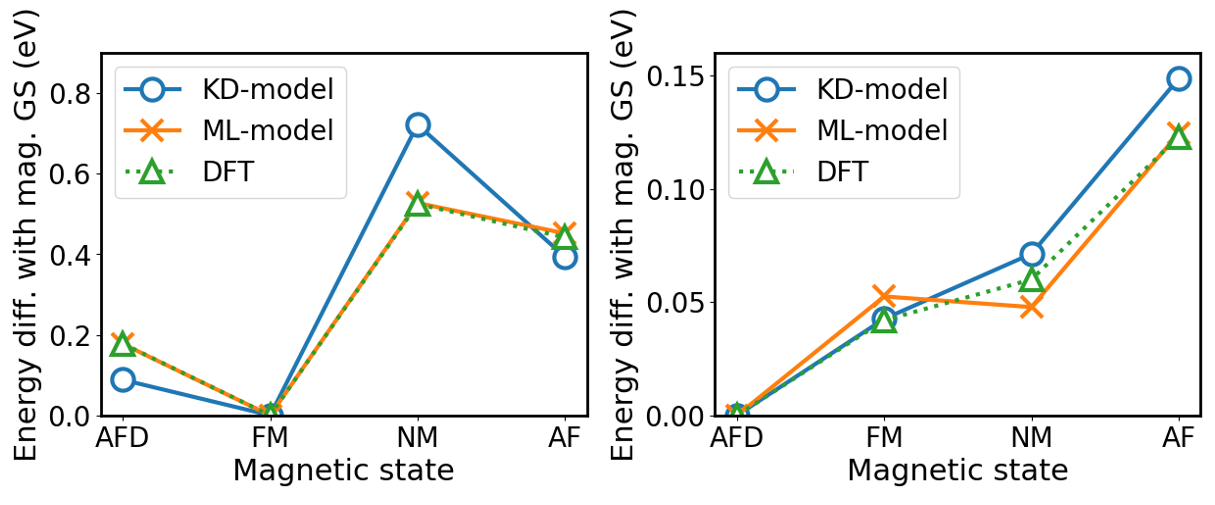}}
  \caption{Model predictions of the energy differences between ordered magnetic states of bcc Fe (left) and bcc Mn (right), using the magnetic moments predicted from DFT, compared to DFT results.  }
  \label{fepur}
\end{figure}

Fig. \ref{fepur} shows the model prediction of the energy difference between the respective magnetic ground-state and various ordered magnetic states of pure bcc Fe and bcc Mn, using
the final sets of parameters. For a close comparison, the  DFT magnetic configurations are used as input for the EIM. The comparison with DFT results shows that the energy hierarchy of the various ordered magnetic states is well reproduced, while some deviations in quantitative energy values (in particular for NM-Fe) result from a compromise of considering various materials properties.

\subsubsection{Fe-Mn magnetic parameters}
\label{model_FeMn}

As for Fe-Fe and Mn-Mn magnetic parameters, the $J_{ij}$s between Fe and Mn spins are also considered up to the $5nn$ shell. They are obtained by fitting to DFT energy differences between Fe-Mn systems (namely isolated Mn solutes, small Mn clusters and Fe-Mn random solutions) with the same atomic configuration but different magnetic structures. 

It is worth mentioning that the presence of a magnetic frustration can be partially resolved by two alternative solutions: either decreasing the spins magnitudes or developing a non-collinear magnetic arrangement. Both features were found in the case of Fe-Cr systems \citep{Soulairol2016}. For the Fe-Mn alloys, the anti-ferromagnetic tendency of Fe-Mn and Mn-Mn interactions, although weaker than the Fe-Cr case, can also induce the emergence of a magnetic frustration at low or intermediate temperatures. 
We note that, using simple local-environment  independent  $J_{ij}$s, many non-collinear ground states are found when performing Monte Carlo simulations. These states are actually an artefact of the EIM, since their energies are significantly higher than those of other collinear states according to our DFT verification.

\begin{figure}[htbp]
  \centerline{\includegraphics[width=1.0\linewidth]{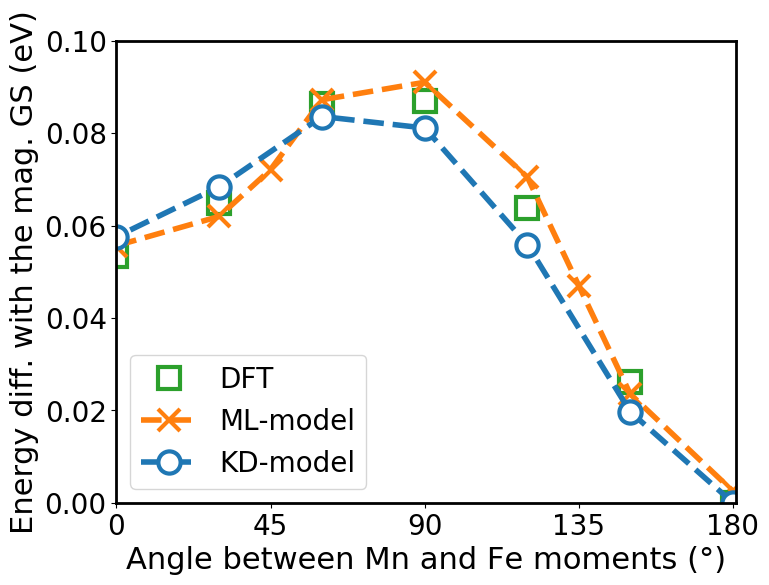}}
  \caption{Energy of an isolated Mn atom in a ferromagnetic Fe matrix with various angles compared to the Fe atoms. The ground-state configuration  (Mn anti-parallel to Fe) is chosen as a reference.}
  \label{noncol}
\end{figure}

In order to solve this problem, we added to the Fe-Mn exchange-coupling parameters a spin-angle dependence, fitted to DFT non-collinear data obtained for an isolated Mn solute in bcc Fe \citep{Schneider2018} (Fig. \ref{noncol}).
The principle is to add a penalty term $J_{0}^{n} \cdot \frac{\theta-90\degree}{90\degree}$ that depends on the angle $\theta$ between the Mn magnetic moment and the average magnetic moment of the Fe atoms in the two nearest-neighbor shells of the concerned Mn atom.
Fig. \ref{noncol} presents the energy dependence of a systems with an isolated Mn atom in a FM Fe matrix on this angle $\theta$. 
As can be observed, our expression allows to reproduce correctly the non-collinear barrier between the two collinear minima with the angle equal to 0\degree (FM-Mn) and 180\degree (AF-Mn).

In addition, as shown by DFT studies Ref.~\onlinecite{Schneider2018}, it is specifically for Fe-Mn alloys very important to consider the dependence of the Fe-Mn magnetic-interaction trend on the Mn concentration in random solutions. In the dilute limit, the Fe-Mn interaction tends to be anti-ferromagnetic, while at higher concentration (above 7 at.\% Mn) it becomes ferromagnetic. In order to properly reproduce this feature, a local concentration dependent term (fourth degree polynomial) is also added to the Fe-Mn $J_{ij}$s.

The final expression for the Fe-Mn exchange-coupling parameters is:
\begin{equation}
\begin{split}
\label{jfemn}
J_{\rm Fe-Mn}^{n} = \left[J_{0}^{n} \cdot \frac{\theta-90\degree}{90\degree}\right] + a\cdot [Mn]_{loc}^{4} \\+ b\cdot [Mn]_{loc}^{3} + c\cdot [Mn]_{loc}^{2}+ d\cdot [Mn]_{loc} +e,
\end{split}
\end{equation}
where the $J_{0}^{n}$ is the original $J_{\rm Fe-Mn}^{n}$ parameter, before considering the angle and concentration dependencies. This parameter ensures the neighbor-shell dependence of the interaction (since the angle and concentration dependencies do not depend on the interatomic distance). 
 $[Mn]_{loc}$ is the local Mn concentration in the five nearest-neighbor shells around the concerned atom.

\subsubsection{Non-magnetic parameters}
\label{model_nomag}

At this point, all the free parameters of the magnetic part of the Hamiltonian are determined and may be used to estimate the magnetic contribution of  the energy difference between two systems. It is for instance possible to calculate the magnetic contribution to the mixing energy of Fe-Mn solid solutions at any concentration, using the following expression:

\begin{equation}
\label{Emixing}
E^{mix}(\text{Fe-Mn}) = \frac{E^{tot}({n\text{Fe}+p\text{Mn}})-nE({\text{Fe}})-pE({\text{Mn}})}{n+p}
\end{equation}

where $E^{tot}({n\text{Fe}+p\text{Mn}})$ is the total energy of the Fe-Mn solid solution, $E({\text{Fe}})$ is the energy per atom of pure bcc Fe (in its lowest energy magnetic state: FM) and $E({\text{Mn}})$ is the energy per atom of pure bcc Mn.

\begin{figure}[htbp]
  \centerline{\includegraphics[width=1.0\linewidth]{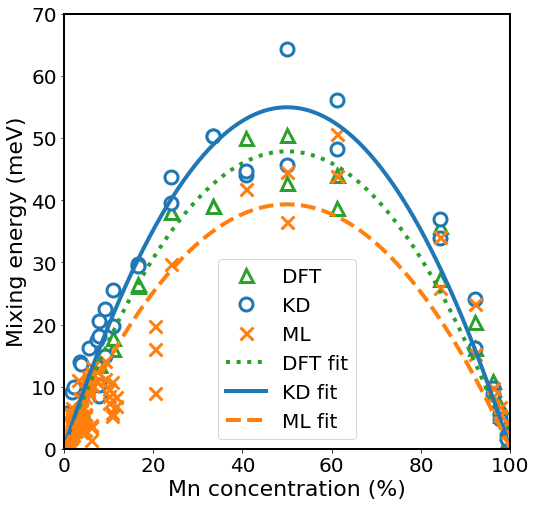}}
  \caption{Total mixing energy obtained from DFT calculations and the effective interaction model. The DFT configurations and magnetic moments have been used as input for the EIM.}
  \label{emixfit}
\end{figure}

The difference between the mixing energy obtained from DFT calculations (which includes magnetic and non-magnetic contributions) and the magnetic contribution of the mixing energy from the EIM provides the non-magnetic contribution of the mixing energy. Chemical parameters of the model are fitted to the latter  in order to accurately reproduce the DFT total mixing energy with the model.
Fig.~\ref{emixfit} shows the mixing energies from DFT and this EIM.  

\subsection{KD model including a vacancy}
\label{model_vac}

With minor modifications, the obtained KD-model can be extended to account for the presence of a small concentration of vacancies, represented by a Fe-Mn supercell containing a vacancy. In this work, we parameterize the model for Fe-Mn alloys that are very dilute in Mn. To do so, we follow a similar approach as described in Ref. \onlinecite{Schneider2020} for the extremely dilute Fe-Cu-vacancy system. We note however that in the present case the effects of vacancies on the local magnetic moment of Mn solutes is more complex than what is observed with Cu solutes. As explained in Ref. \onlinecite{Schneider2018}, the presence of a vacancy highly favors the AF-Mn state to the detriment of FM-Mn. Such complexities have to be taken into account when parameterizing the interactions in the presence of a vacancy. Also, in Ref. \onlinecite{Schneider2020} the model was parameterized for the extremely dilute case of one Cu solute and one vacancy in bcc Fe, while in the present case, the model was first parameterized without vacancies for various concentrations (see Section \ref{model I}). Only then the parameters were adjusted to take into account the presence of a vacancy nearby a Mn solute, which makes the approach slightly different.

The overall principle is to preserve the parameterization obtained for the defect-free Fe-Mn system, and to include some variations in the parameters for atoms near the vacancy. 
The on-site $A_{i}$ and $B_{i}$ parameters of both Fe and Mn species are modified for atoms located at the first and second nearest-neighbor ($1nn$ and $2nn$) shells of the vacancy in order to model the magnitude variation of the magnetic moments induced by the presence of a vacancy. Indeed, DFT results show that nearby a vacancy, the $1nn$ (resp. $2nn$) Fe atoms magnetic moment magnitude tend to increase (resp. decrease) by 0.2 $\mu_{B}$ \citep{Schneider2020}.
Also, $J_{Fe-Mn}$ parameters are modified for the atoms at $1nn$ and $2nn$ sites of the vacancy in order to capture the change of the relative energetic stability of the two magnetic minima of a Mn atom in Fe. For example, as predicted by DFT calculations\cite{Schneider2018}, for an isolated Mn in Fe, the state with the Mn spin anti-parallel to the Fe spins (AF-Mn) is 0.05 eV lower in energy than the state with the Mn spin parallel to the Fe spins (FM-Mn). But, if the Mn solute is at $1nn$ of a vacancy, this energy difference increases to 0.25 eV.

\subsection{Machine-learning (ML) model}
\label{model II}

For the parameterization of the machine-learning model, a ridge regression approach was employed. This method has previously been used to successfully predict interatomic interactions in non-magnetic elemental metals after a training with thousands of DFT data \cite{seko2019group,takahashi2017conceptual,takahashi2018linearized}. Recently, a similar type of regularization was employed to obtain a Heisenberg spin Hamiltonian \cite{li2020constructing}. Such a regression allows us to obtain a parameter set without evaluating the DFT data points individually, while the regularization is important to inhibit the problem of over-fitting. However, for the sake of consistency with the knowledge-driven model, the prediction only accounts for the same energy values and ignores their derivatives and forces. As one of the consequences, it does not necessarily reproduce the correct spin structure for the ground state in the Mn-rich regime.

The regression is performed for the complete training set of collinear DFT data, but the selection of a proper starting point is subject to the two-step approach. Accordingly, the model parameters for pure Fe were fitted firstly to the DFT data containing only Fe atoms in collinear calculations. The parameters obtained in this step are then used as input parameters for the fitting of all model parameters in the second step. This two-step approach exploits the multivariate normal distribution of the prior probability distribution of the fit parameters following the Bayesian interpretation of the ridge regression \cite{vogel2002computational}. 
In other words, the probability of finding the fit parameters for pure Fe in the second fitting step is given by normal distributions, for which the variance is given by the regularization factor. Finally, the non-collinear terms were determined separately via a standard least square method. 
For the ridge regression method, the data points were separated into 10 training sets, out of which one set was used for validation  at each cross-validation step. 
The machine-learning model uses the same model Hamiltonian (\ref{total_hamil}) as before with minor modifications as explained below.

\subsubsection{Magnetic parameters for pure bcc Fe and Mn}

In order to meet the above-mentioned challenges to reproduce the magnetic ground state in the Mn-rich regime, the on-site terms for the Mn magnetic moment magnitude involve $A_{\rm Mn}$, $B_{\rm Mn}$ and $C_{\rm Mn}$ terms (see Eq.~(\ref{total_hamil})) whereas in the case of Fe the description is limited to the two on-site terms $A_{\rm Fe}$ and $B_{\rm Fe}$.

Fig. \ref{abcomp} shows the variation of energy per atom for the on-site terms as a function of the magnetic moment. As compared to the KD model, the stabilization of the Fe moment is slightly more pronounced, yielding almost the same moment magnitude. For the Mn atoms, however, the magnetic moments are even more dispersed than in the KD model, with hardly any change in the energy until the magnitude becomes relatively large ($|M|>3$).

The two-step approach ensures that even after the regression with all collinear DFT results the ground states of pure Fe can be reproduced with an error per atom around 1~meV (\emph{cf.} fig.~\ref{fepur}). We can see that the energy values for pure Mn were not reproduced with the same precision. This is also because the dataset mostly consists of Fe-rich DFT results, so that for the sake of reproducing Mn interactions embedded in Fe, the pure Mn interactions had to be compromised. As a consequence, the machine-learning model puts more emphasis on pure Fe than the KD approach, which yields the better description of pure FM Fe.

\subsubsection{Fe-Mn pair-interaction parameters}

The Fe-Mn Heisenberg parameters in the machine-learning model contain 5 different values for the 5$nn$ shells, \emph{i.e.} have the same form as for pure Fe and pure Mn. However, a correction term $c_{\rm Mn}\Delta J_{\rm Fe-Mn}$, which depends on the global Mn fraction in the system $c_{\rm Mn}$, was added to each of the 5 Heisenberg parameters. This is different to Eq.~(\ref{jfemn}), which explicitely contains the local Mn concentration.
Furthermore, an angle-dependent contribution $J_0^n|{\bf M}_{\rm Mn}\times \langle {\bf M}_{\rm Fe}\rangle|^2/|{\bf M}_{\rm Mn}|^2$, where $\langle{\bf M}_{\rm Fe}\rangle$ is the average magnetic moment of Fe, was added to each Mn atom.

Fig.~\ref{noncol} shows the variation of the angle-dependent part for the Fe-Mn interactions. Similarly to the knowledge-driven model, it reproduces the non-collinear barrier from AF to FM spin structures as manifested by the DFT results.

For consistency with the KD-model, Non-magnetic interactions are considered up to the fifth nearest-neighbor in the case of Mn-Mn interactions, and up to the second nearest-neighbor for Fe-Mn and Fe-Fe interactions. Contrary to the fitting procedure of the KD-model, these parameters are fitted altogether with the magnetic ones, by taking into account the differences of total energy between the various configurations obtained via DFT calculations.

Together, the ML model contains fewer analytical terms for the composition dependence of the exchange-coupling parameters, consistent with the philosophy of an automatized machine-learning approach.

We note that values of the model parameters obtained with the two different fitting techniques largely differ in some cases by several orders of magnitude (e.g., $V_{\rm Fe-Fe}$) or even the sign (e.g., $J^{(3nn)}_{\rm Fe-Fe}$). This underlines the completely different fitting strategy of both approaches.

\section{Ground-state properties: Accuracy of the models}
\label{validation}

In this section,  the accuracy of the two models is verified and discussed through a comparison with DFT results on properties of bcc Fe, bcc Mn and bcc Fe-Mn systems. In each case, equivalent atomic configurations as in the DFT data are used, whereas the magnetic structures for the models are determined using Monte Carlo spin relaxations.

\begin{figure}[htbp]
  \centerline{\includegraphics[width=1.0\linewidth]{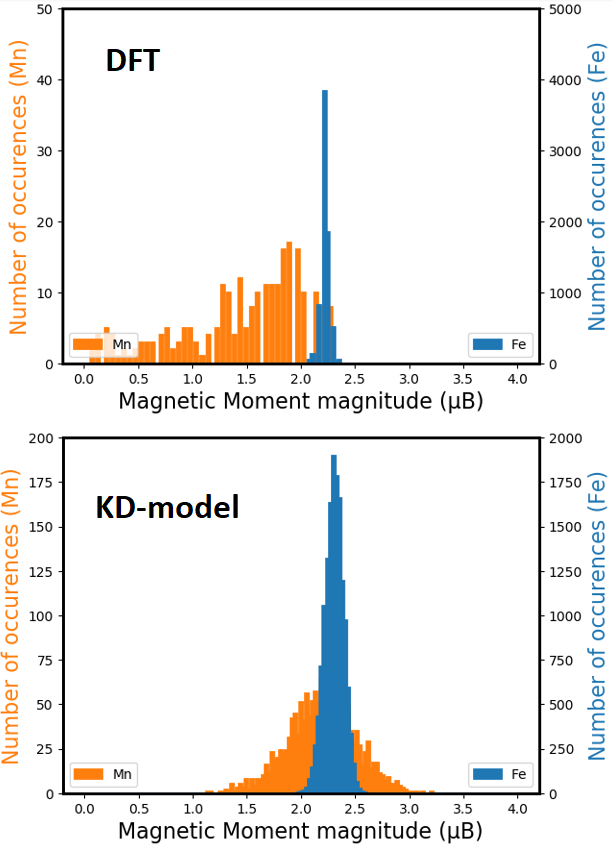}}
  \caption{Distribution of Fe and Mn local magnetic moments in Fe-Mn SQS alloys, up to 50 at.\% Mn, using (top) DFT calculations and (bottom) Interaction model coupled with Monte Carlo simulations at .}
  \label{distrib}
\end{figure}

First, the distribution of Mn and Fe magnetic moments magnitudes in random solutions up to 50 at.\% Mn from DFT calculations\cite{Schneider2018} is shown in the top panel of Fig. \ref{distrib}. The data reveal the complexity of Mn magnetism in bcc Fe, with very scattered magnetic moment magnitudes. The bottom panel of Fig. \ref{distrib} shows the corresponding distribution obtained using the KD-model along with spin Monte Carlo simulations at . Concerning the distribution of Fe magnetic moment magnitudes, the agreement between the two approaches is excellent. In the case of Mn, although it is difficult to model such a complex behavior with a simple model, both approaches show the same general trend: a maximum around 2.0 $\mu_{B}$, and a wider distribution than Fe.

Concerning the pure bcc Mn, when using the DFT predicted magnetic state as an input, the AFD state is properly predicted (with both KD and ML models) as the lowest energy magnetic state, compared to FM, AF and NM states (see Fig. \ref{fepur}). However, when performing spin-MC simulations, both models cannot capture the direction-dependent magnetic interaction. Using the KD-model, the magnetic ground-state predicted by Monte Carlo simulation is a spin-glass without any magnetic long-range order, which energy is 0.02 eV/atom lower than the AFD ground-state.
Due to the limited number of DFT training data, this applies even stronger to the machine-learning model, which also shows a disordered magnetic state, which energy is 0.8 eV/atom lower than the AFD ground-state.

\begin{figure}[htbp]
  \centerline{\includegraphics[width=1.0\linewidth]{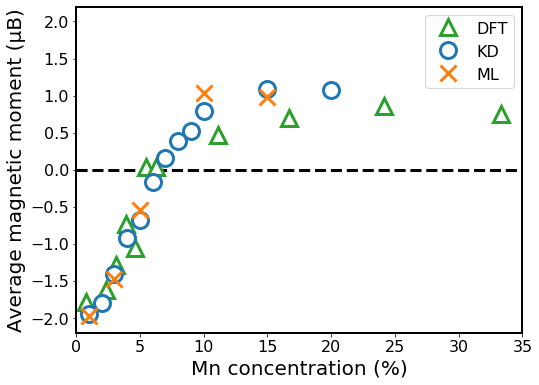}}
  \caption{Average magnetic moment of Mn atoms as a function of Mn concentration, from DFT calculations and spin Monte Carlo simulations at .}
  \label{transitionSQS}
\end{figure}

\begin{figure}[htbp]
  \centerline{\includegraphics[width=1.0\linewidth]{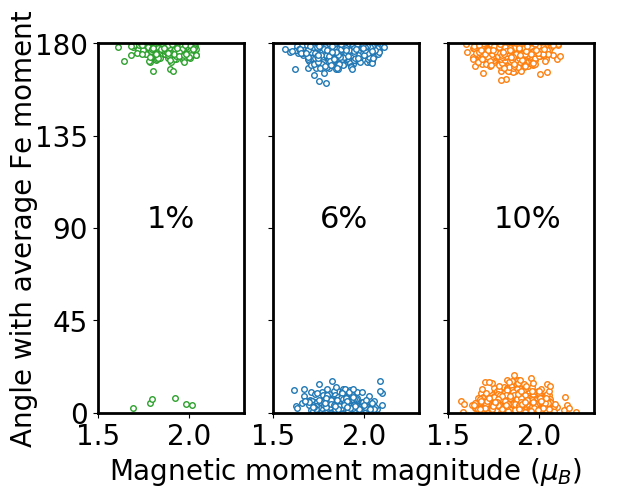}}
  \caption{Angle distribution between Mn magnetic moments and the average magnetic moment of Fe atoms, in random solutions at 1, 6 and 10 at. \% Mn concentrations, using the KD-model along with spin Monte Carlo simulations at .}
  \label{repartition}
\end{figure}

Both EIMs allows us to properly simulate the concentration dependency of the Fe-Mn magnetic interaction tendency, as shown in Fig. \ref{transitionSQS}, in comparison with DFT results. The change of average Mn magnetic state is also explicited by the angle distribution of Mn magnetic moments compared to the average magnetic moment of Fe atoms. The results shown in Fig. \ref{repartition} are obtained with spin Monte Carlo simulation at 10K in random Fe-Mn solutions at 1, 6 and 10 at. \% Mn, using the KD-model. It is clear that at 1\% the coupling tendency between Mn and Fe moments is anti-ferromagnetic while increasing the concentration favors more and more the ferromagnetic coupling. Also, in agreement with DFT predictions, non-collinear states are not predicted for such a low temperature.

As explained in Sec. \ref{eim}, the mixing energy of the Fe-Mn random solutions was used to parameterize the interaction parameters of the models. At that stage, the model were shown to correctly predict the mixing energies when adopting the magnetic moments of the DFT data (Fig. \ref{emixfit}). Here, we investigate whether the EIMs can satisfactory predict mixing energies at their own magnetic ground states.

The concentration dependence of bcc Fe-Mn mixing energy is determined by generating Fe-Mn random alloys at various concentrations. The magnetic state of these random configurations is relaxed via spin Monte Carlo simulation at 1 K, while the atomic structure is kept constant in order to prevent the possible appearance of atomic short range order or any phase separation.

\begin{figure}[htbp]
  \centerline{\includegraphics[width=1.0\linewidth]{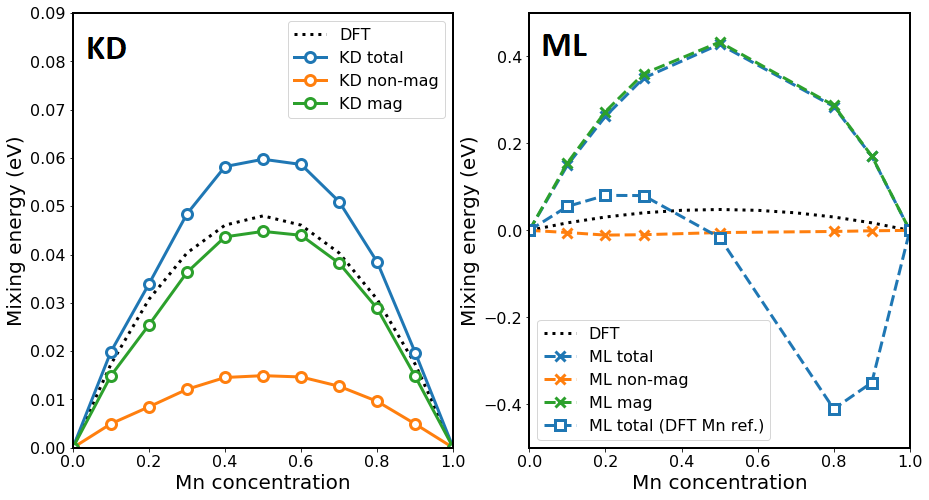}}
  \caption{Mixing energy of random solutions as a function of Mn atomic concentration. The black dotted line shows the results obtained with DFT for comparison, the blue line shows the total mixing energy obtained from T=1K spin-MC simulations, the green and the orange lines respectively show the magnetic and the non-magnetic contributions of the mixing energy obtained via MC simulations.}
  \label{0K_mixing}
\end{figure}

Within the KD model the bcc Mn reference state presents a spin glass as the lowest energy magnetic state, with an energy close to that of the DFT magnetic ground-state. In case of the ML model, however, pure Mn and the Mn-rich region are less accurately captured (see  Sec.~\ref{model II}). Therefore, in the latter case, we provide in Fig. \ref{0K_mixing} the mixing energy curves considering both the DFT and the ML-model magnetic ground-states.

As can be seen in Fig. \ref{0K_mixing}, the obtained mixing energies are positive for all concentrations with the KD-model and are very close to the DFT data. When considering the ML-model with the MC relaxed magnetic state for the pure Mn reference, all the mixing energies are positive although they exhibit too large values (about ten times larger than the KD-model and the DFT data). When considering the ML-model with the DFT magnetic state for the pure Mn reference instead, the mixing energy is positive only up to 50\% Mn. Accordingly the agreement with the present and previous\cite{Lintzen2013} DFT results differs for both models. In particular, the results for the KD-model indicate an unmixing tendency that is consistent with  experimental evidences \cite{Hillert1977}.

A qualitative difference between the two models can be observed when separating the magnetic and non-magnetic contributions of the mixing energy,  as shown in Fig. \ref{0K_mixing}. Using the KD-model, both magnetic and non-magnetic contributions exhibit positive values with a similar order of magnitude (although the magnetic contribution shows larger values). Using the ML model, where both contributions are fitted simultaneously, the magnetic contribution are extremely dominant, whereas the non-magnetic terms are negative with very small values.

This discrepancy between the two models illustrates that the number of DFT training data and the available information about the stability of input structures are insufficient to accurately describe mixing in an automized ML approach. To support the fact that artificial intelligence based approaches generally require extended datasets, the example of previous studies in Refs. \onlinecite{Soisson2007} and \onlinecite{Messina2017} can be mentioned. These two studies lead to very similar conclusions concerning the precipitation kinetics in Fe-Cu bcc alloys, although the knowledge-driven one \citep{Soisson2007} considers a few tens of barriers while the machine-learning based study \citep{Messina2017} requires 2000 barriers.
At the same time, our results also indicate that the energetic properties are much more sensitive to the parameterization than the magnetic properties. The prediction of the latter is visibly more robust.

\begin{figure}[htbp]
  \centerline{\includegraphics[width=1.0\linewidth]{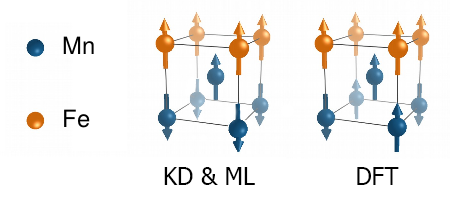}}
  \caption{Visualization of the magnetic ground-state of the 5-Mn cluster as predicted using the models (both KD and ML) and DFT.}
  \label{clusters}
\end{figure}

Besides the magnetic properties of the Fe-Mn random solutions, we have also verified the prediction of magnetic ground state of Mn clusters in bcc Fe, in view of the unmixing tendency of the alloy.
We find that the two models correctly predict the DFT ground-state of every Mn-cluster configuration from 2 to 8 atoms (see Fig. 10 of Ref. \onlinecite{Schneider2018}), except in the case of the 5-Mn cluster (see Fig. \ref{clusters}) where the ground-state obtained from the EIM-MC simulation (using both the KD and the ML models) is found to be 0.01 eV/Mn less energetic than the ground state predicted by DFT calculations.

\smallbreak
\section{Model predictions of temperature-dependent properties}
\label{results}

In this section, we provide examples to illustrate the applicability and accuracy of the present models, for the prediction of finite-temperature properties. We present the results of the KD-model for all the properties. When relevant, the outcomes from the two models are compared. Such comparisons allow us to assess the robustness and quality of the predictions in scenarios for which DFT calculations are not feasible.
The first four subsections address magnetic and thermodynamic properties at fixed atomic configurations, via spin Monte Carlo equilibrations, while the last two subsections investigate the interplay between magnetic and chemical configurations, employing a coupled spin-atom MC simulations.

\subsection{Temperature dependence of bcc Fe magnetic properties}
\label{pure_Fe_T}

\begin{figure}[htbp]
  \centerline{\includegraphics[width=1.0\linewidth]{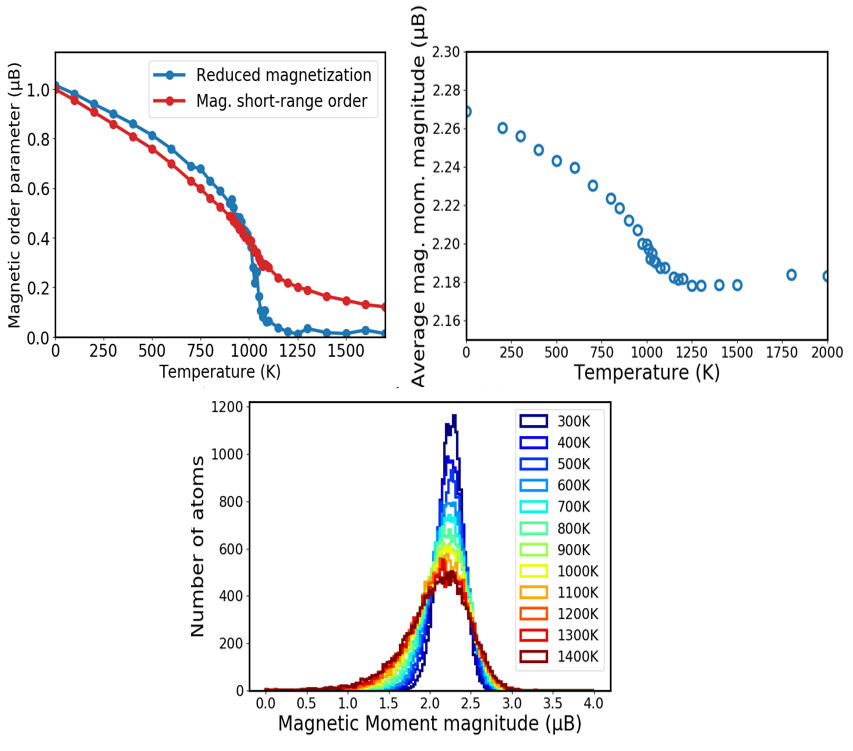}}
  \caption{(top left) Temperature evolution of the reduced magnetization and the $1nn$ magnetic short-range order in pure bcc Fe. (top right) Temperature evolution of the average magnetic moment magnitude of Fe atoms. (bottom) Distribution of the magnetic moment magnitude of Fe atoms for various temperatures. The three panels are produced using the KD-model along with spin Monte Carlo simulations.}
  \label{magnetization}
\end{figure}

First of all, the reduced magnetization of pure Fe is shown in Fig. \ref{magnetization}. At each temperature, it is obtained by averaging the magnetic moments over the whole system and normalizing the numbers by the corresponding averaged magnitude of the magnetic moments. The calculated $T_{C}$ is approximately 1060K, compared to 1044K obtained experimentally \citep{Keffer1966}. The result is almost the same for both models, indicating that the slight deviations for AFD and NM energies in the KD model (Fig.~\ref{fepur}) have no impact on this property.

The $1nn$ magnetic short-range order (MSRO), defined here as the nearest-neighbor spin pair-correlation function is also shown in Fig. \ref{magnetization}. As can be noticed, in the low temperature domain (below $T_{C}$) it decreases with temperature slightly faster than the magnetization. On the opposite, at high temperature (beyond $T_{C}$), MSRO remains significant . These results are in good agreement with previous studies \citep{Haines1985, Lavrentiev2009, Melnikov2019}. The significance of MSRO in both models, with only slightly larger values in the machine-learning model, indicates the robustness and physical relevance of this prediction. Therefore, the study of properties around $T_{C}$ needs to take MSRO into account.

As shown in the upper panel of Fig. \ref{magnetization}, the average magnitude of Fe magnetic moments decreases with temperature up to the Curie temperature. When $T>T_{C}$, the average magnitude increases very slightly with temperature. This curve is in good agreement with the results of Lavrentiev {\it et al.} \citep{Lavrentiev2010} obtained with a similar approach. However, one should note that these variations of the average magnitude are very small (contained within 0.1 $\mu_{B}$) which suggests that the classical Heisenberg model is a good approximation for pure Fe.
 
The resulting temperature evolution of the magnitude distribution, shown in the bottom panel of Fig. \ref{magnetization}, is a direct consequence of the magnetic on-site terms shown in Fig.~\ref{abcomp}. The results are in good agreement with a study of Ruban {\it et al.} \citep{Ruban2012}, performed with a similar approach.

As explained in the previous section, the analysis of the KD and ML models results shows that the magnetic ground-state predicted by Monte Carlo simulation is a spin-glass without any magnetic long-range order. Because of this, a N\'eel transition, which might occur in bcc Mn and in the Mn-rich limit of the alloy, is not reproduced. There is no experimental evidence of such a N\'eel transition because the bcc phase of pure Mn is only stable at very high temperature (between 1411 and 1519 K). However, as we have shown in a previous study that the DFT ground-state of pure bcc Mn is AFD \citep{Schneider2018}, we expect a magnetic transition, going from this state to the PM state.
As we are mostly interested in the Fe-rich part, we believe it is not crucial here to properly describe such properties in the extremely Mn-rich domain.

\subsection{Curie temperature of bcc Fe-Mn random solutions}
\label{Tc}

The Curie temperature is a fundamental property of ferromagnetic systems. As our goal is to develop an effective interaction model, capable to describe properly the magneto-thermodynamic properties of the bcc Fe-Mn alloys at any given temperature, the Mn concentration dependence of $T_{C}$ in the dilute Fe-Mn alloys is of great relevance.

\begin{figure}[htbp]
  \centerline{\includegraphics[width=1.0\linewidth]{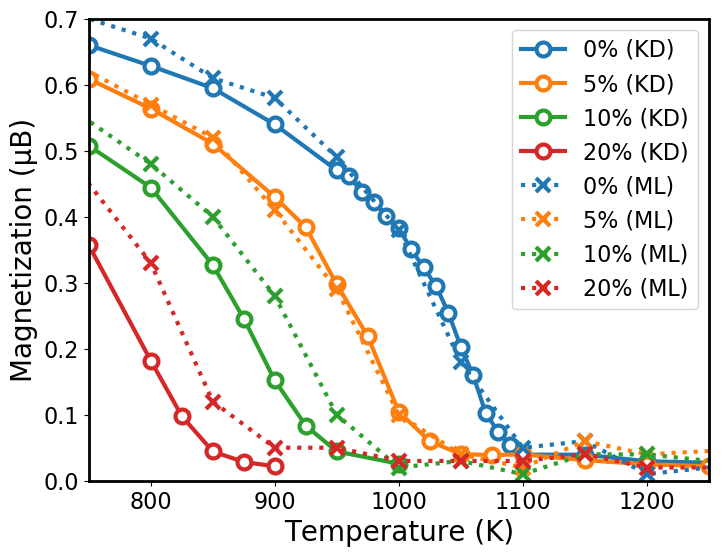}}
  \caption{Temperature evolution of the total magnetization, for various Mn atomic concentrations in the Fe-rich regime. The results of the KD model (solid lines) and the ML model (dashed lines) are compared.}
  \label{MT_mn}
\end{figure}

In Fig. \ref{MT_mn}  the temperature dependence of total magnetization for various Mn concentrations in the Fe-rich regime is provided. These calculations are all performed in random solutions. The magnetization curves are systematically shifted towards low temperatures with increasing Mn concentrations. For dilute bcc Fe-Mn alloys a remarkable agreement between both models is observed, despite the fact that even small differences in the analytical expressions of the Fe-Mn exchange-coupling parameters are used.  
For 10 and 20 at.\% of Mn the shift is slightly stronger in the KD model as compared to the ML model.

\begin{figure}[htbp]
  \centerline{\includegraphics[width=1.0\linewidth]{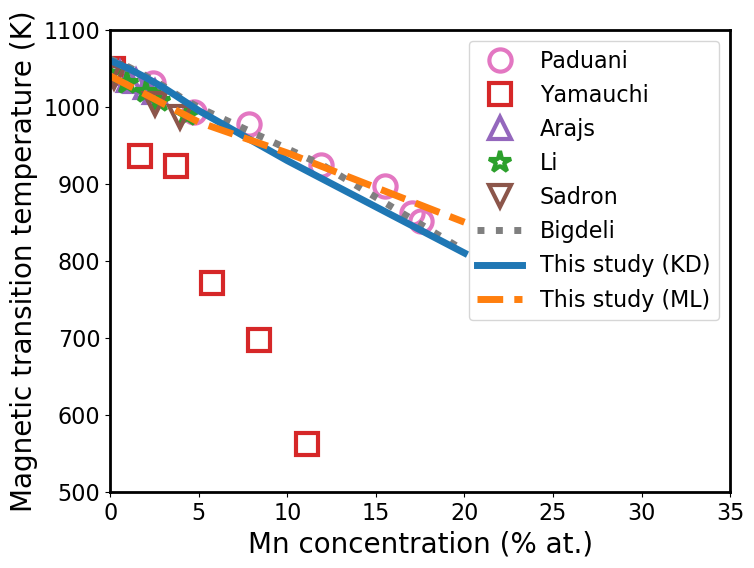}}
  \caption{Curie temperature of dilute bcc Fe-Mn alloys as a function of Mn concentration. The solid and dashed blue lines show our results while the symbols show literature experimental results. circles: Paduani {\it et al.} \citep{Paduani1991}, squares: Yamauchi {\it et al.} \citep{Yamauchi1974}, upward triangles: Arajs {\it et al.} \citep{Arajs1965}, stars: Li {\it et al.} \citep{Li2002}, downward triangles: Sadron {\it et al.} \citep{Sadron1932}. Our results are also compared to a CALPHAD study shown with a dotted line: Bigdeli {\it et al.} \citep{Bigdeli2019}}
  \label{tc_mn}
\end{figure}

We estimate the Curie temperature $T_{C}$ as the inflection point of the $M(T)$ curve\citep{Fabian2013}. 
The $T_{C}$ of each considered concentration is reported in Fig. \ref{tc_mn} in order to compare with existing experimental results. As can be seen, $T_{C}$ decreases with Mn concentration with a slope of approximately 10K per Mn at.$\%$, in excellent agreement with most literature data. Indeed, most experimental works have shown that $T_{C}$ tends to decrease in the dilute limit linearly with Mn concentration, at a rate of approximately 10K per Mn at.\%  \citep{Paduani1991, Arajs1965, Li2002, Sadron1932}, as shown in Fig. \ref{tc_mn}. 

For intermediate compositions up to 20 at.\% of Mn, deviating experimental trends are reported in the literature. While the slope of 10 K per Mn at.\% continuous in the case of Paduani {\it et al.} \citep{Paduani1991}, Yamauchi {\it et al.} \citep{Yamauchi1974} found a larger decreasing slope of around 43K per Mn at.\%. 

One possible explanation for the deviation of Yamauchi {\it et al.} is that their magnetic measurements are biased by the use of cold-rolling on the samples in order to stabilize the body-centered cubic phase \citep{Yamauchi1974}, which is not the case in the other experimental studies. A recent study indeed suggests that a plastic deformation has a significant impact on the atomic short-range order of $\alpha$-Fe-Mn systems \citep{Shabashov2017}, which according to our results may affect the magnetic configuration. We note that in order to stabilize $\alpha$-Fe-Mn beyond 5 at. \%  Mn, Paduani {\it et al.} have added 3 at. \% Ti to the solution \citep{Paduani1991}. As the low concentration results (below 5 at.\%  Mn) of Paduani {\it et al.} are in excellent agreement with the studies using pure Fe-Mn, it can be assumed that there is not significant effect of such Ti addition on the magnetic state of the solution. A recent Calphad assessment \citep{Bigdeli2019} also assumes such a decrease of $T_{C}$ with Mn concentration at a rate of approximately 10K per Mn at.\%, as shown in Fig. \ref{tc_mn}.

\smallbreak

\subsection{Temperature dependence of Mn magnetic moment}
\label{magmn_temp}

As shown in Fig. \ref{transitionSQS}, the average magnetic moment of Mn atoms in bcc Fe-Mn solid solutions obtained from both EIMs at very low temperature (1K) Monte Carlo simulations shows the same Mn concentration dependence as predicted by DFT calculations. It tends to be anti-ferromagnetic to Fe magnetic moments at low concentration (below the transition, which occurs at about 7 at.\%  Mn) and ferromagnetic at high concentrations.

Using our models with spin Monte Carlo simulations, we determine the evolution of the average magnetic moment of Mn atoms with temperature, at fixed random atomic configurations in order to avoid the appearance of atomic short-range order. Fe-Mn random alloys at 0.1,  1 and 10 at.\%  Mn are studied, in order to consider both Mn magnetic regimes, below and beyond the Mn magnetic-state transition concentration.

\begin{figure}[htbp]
  \centerline{\includegraphics[width=1.0\linewidth]{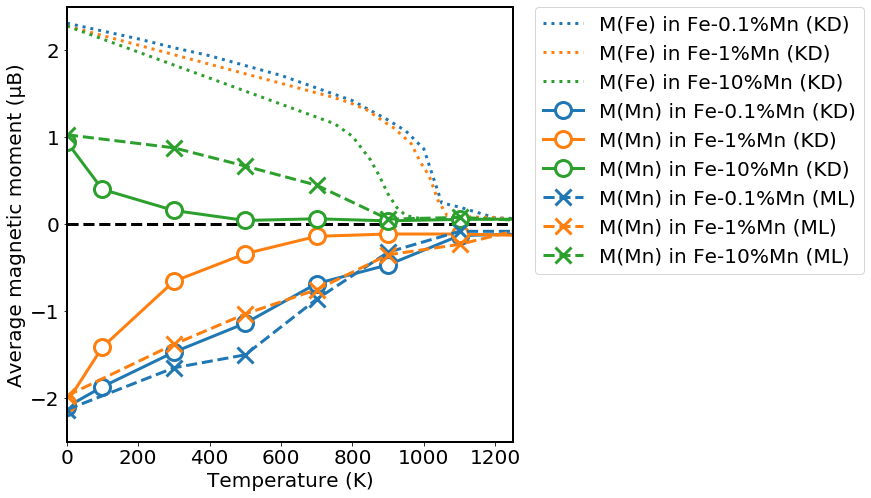}}
  \caption{Temperature evolution of the average magnetic moment of Fe and Mn atoms in bcc Fe-Mn at various concentrations (0.1, 1 and 10 \% at. Mn). The total magnetization of pure bcc Fe is shown as a black dotted line for shape comparison.}
  \label{mnmagt}
\end{figure}

As can be observed in Fig. \ref{mnmagt}, in both regimes and for both models the absolute value of the average magnetic moment of Mn atoms decreases with temperature faster than the average moment of Fe (which decreases following the evolution of the total magnetization). In the truly isolated-Mn case (0.1 at.\% Mn) the decrease is almost linear and the curve reaches 0 approximately at the Curie temperature of the system. 
For the more concentrated cases, the two parameterizations yield qualitatively different trends. While the ML model predicts an almost linear of the Mn magnetic moment below and above the magnetic-state transition concentration, the decrease is faster in the case of the KD model.

Additional analysis of our data confirms that the magnetic moment magnitude of Mn atoms tends to increase with temperature, from around 1.85 to 2.05 $\mu_{B}$ in the considered temperature range. This attests that the loss of average magnetic moment does not come from the longitudinal spin excitations.
A plausible cause of this fast decrease of the average Mn moment magnitude compared to the Fe case is the atypical presence of two magnetic minima for a Mn atom in Fe, namely the AF- and the FM-states, with a rather small energy difference (0.05 eV for an isolated Mn at 0K). The AF state is the ground state for the isolated Mn, but the FM state becomes gradually populated with increasing temperatures. 
     
For a comparison, we consider the same magnitudes in dilute bcc Fe-Cr alloys, where Cr magnetic moments always  tend to be anti-parallel to Fe moments \citep{Trochet2018}. These results  however show that the Cr average moment follows the same decrease shape as in the Fe case, and at variance with the Mn case.

\begin{figure}[htbp]
  \centerline{\includegraphics[width=1.0\linewidth]{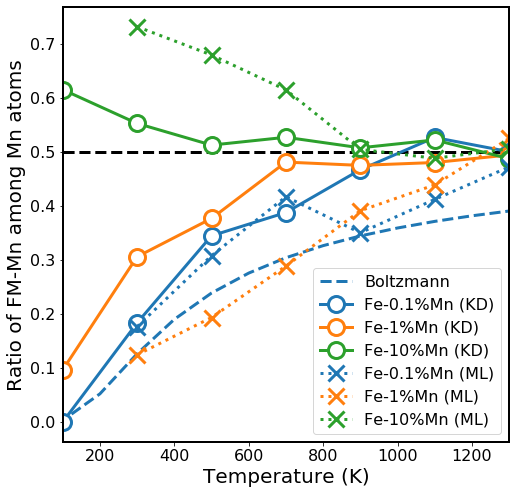}}
  \caption{Temperature dependence of the ratio of FM-Mn among Mn atoms in bcc Fe-Mn at various concentrations (0.1, 1 and 10 \% at. Mn). For the case of 0.1 at.\% Mn the expected ratio from Boltzmann theory (see text) is shown with the dotted line.}
  \label{mn_ratio}
\end{figure}

In order to go further in the analysis, we have determined the ratio of Mn atoms at the FM state (FM-Mn) 
as a function of the temperature. The results are shown in Fig. \ref{mn_ratio}, We can indeed observe that, in agreement with DFT data, the ratio of FM-Mn atoms at low temperatures is approximately 0\% for the 0.1 and 1 at.\% Mn concentrations, and is above 60\% for the 10 at.\% Mn case. As temperature increases, the ratio of FM-Mn evolves towards 50\%. The details of the temperature dependence, however, is qualitatively different in both parameterizations, what explains the different behaviour in Fig.~\ref{mnmagt}.

We also compared in Fig.\ref{mn_ratio} the ratio of FM-Mn in the 0.1 at \% Mn system from our Monte Carlo simulations and the expected ratio from the Boltzmann theory, expressed as follows:

\begin{equation}
\label{boltzmann}
\frac{N_{FM-Mn}}{N_{AF-Mn}+N_{FM-Mn}} = \frac{\exp(\frac{-\Delta E}{k_{B}T})}{1+\exp(\frac{-\Delta E}{k_{B}T})}
\end{equation}
with $N_{FM-Mn}$ and $N_{AF-Mn}$ being respectively the number of FM-Mn and AF-Mn atoms, and $\Delta E$ being the energy difference between FM-Mn and AF-Mn states obtained from DFT calculations ($\Delta E$ = 0.05 eV at 0k predicted by DFT and both models).

We note that as temperature increases, the Fe magnetic state becomes more and more disordered, and the terms AF-Mn and FM-Mn are less and less defined. Especially, at temperatures above the Curie point, as the system is paramagnetic, there is no reason to expect the ratio to follow properly the Boltzmann distribution. We keep classifying Mn atoms depending on the direction of their spins along an arbitrary axis in order to illustrate their random distribution at high temperatures.

\smallbreak

\subsection{Temperature dependence of Fe-Mn mixing energy}
\label{mixing_temp}

In the previous section, the mixing energy of bcc Fe-Mn random alloys is calculated over the whole range of concentrations at 1K, using Monte Carlo simulations. Similar spin-MC simulations were performed at various temperatures in order to study the effect of temperature-dependent magnetism on the mixing energy of bcc Fe-Mn alloys.

\begin{figure}[htbp]
  \centerline{\includegraphics[width=1.0\linewidth]{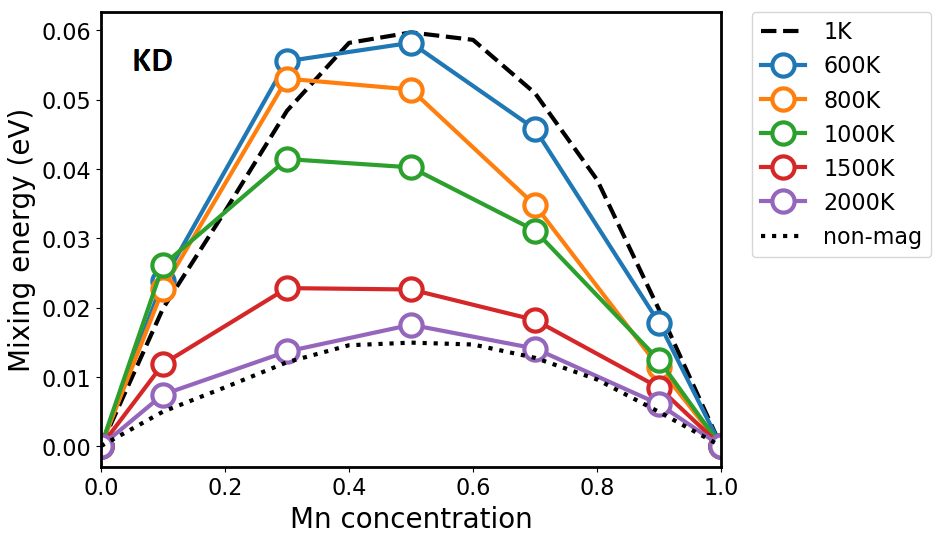}}
  \caption{Mixing energy of random Fe-Mn alloys as a function of Mn atomic concentration for various temperatures. The black dashed line shows the low temperature limit of mixing energy while the black dotted line shows the non-magnetic contribution, also corresponding to the high temperature limit.}
  \label{emix_T}
\end{figure}

As shown in Fig. \ref{emix_T}, the temperature evolution of the bcc Fe-Mn mixing energy is related to the magnetization. Indeed, for any given concentration, two regimes can be clearly identified: below the magnetic transition temperature, the mixing energy curve remains quite similar to the 0K limit, while beyond the magnetic transition temperature, it converges to a high temperature paramagnetic limit. It can also be noticed that at both high and low temperature limits, that is, as soon as for all the concentrations the temperature is located either below or above the magnetic transition temperatures, the mixing-energy curve is symmetric. But, for the intermediate temperatures, an asymmetry appears between the Mn-rich and the Fe-rich domains. Also, the Mn-rich mixing energies decrease faster with increasing temperature than the Fe-rich values. This asymmetry is consistent with the above-mentioned two-regime behavior, considering the difference between the Fe-rich and Mn-rich magnetic transition temperatures. As previously described, we predict the magnetic transition temperature to decrease with Mn concentration (see Fig. \ref{tc_mn}). Actually, the Mn-rich phase should exhibit a Neel magnetic transition temperature regarding the DFT magnetic ground-state, which we cannot reproduce as our model predicts a spin-glass magnetic structure at low temperature in the Mn-rich side.

Please note that the decrease of the mixing energy with temperature is only due to the magnetic contribution as there is no atomic-position changes in these simulations. We notice that the paramagnetic limit (2000 k) of the mixing energy curve is very similar to the non-magnetic contribution of the 0K mixing energy (shown in Fig. \ref{emix_T}), indicating that the mixing between Fe and Mn atoms has a negligible impact on the average magnitude of their respective magnetic moments. Overall, the present results suggest that spin disordering favors the mixing of Fe and Mn, while spin ordering favors the phase separation tendency.

It is worth mentioning that our results are in qualitative agreement with a previous CALPHAD prediction \citep{Bigdeli2019}, in which the thermodynamic parameters lead to a fully positive mixing energy of bcc Fe-Mn alloys which decreases with temperature. For the sake of comparison, the mixing energy at 50\% at. Mn calculated using the parameters of this study is 0.08 eV at a 1K temperature (our value being 0.06 eV). Concerning the decrease rate, the KD-model predicts that the 50\% at. Mn mixing energy converges to 0.015 eV around 2000K while the mixing energy calculated using the parameters of Ref. \onlinecite{Bigdeli2019} shows a slower decrease (0.06 eV at 2000K).

As we have shown in the previous section that the low temperature mixing energy predictions of the ML-model exhibit qualitative differences with DFT results, we chose not to develop the temperature dependence of this property using the ML-model.

\smallbreak

\subsection{Temperature and concentration evolution of atomic short-range-order}
\label{unmixing}

The Monte Carlo results presented up to this point are performed by varying only the magnetic configuration of the system while the atomic structure is frozen. In order to go further insights into the interplay between the magnetism and thermodynamic properties versus temperature, it is necessary to follow the evolution of both the magnetic and the atomic structures simultaneously. Therefore, we include, in addition to the Monte Carlo spin equilibration,  atomic exchanges on a bcc lattice. 

To this end, bcc Fe-Mn alloys were studied at various concentrations and temperatures, in order to evaluate the clustering tendency for both degrees of freedom. 
We consider the Cowley-Warren formulation of atomic short-range order (ASRO) \citep{Cowley1950,Warren1990}, for which the parameter
\begin{equation}
\label{cw_sro}
\alpha_{i}^{\rm Mn} = 1 - \frac{n_{i}}{Z_{i}C_{\rm Fe}}
\end{equation}
is averaged over all Mn atoms. Here, $n_{i}$ is the number of Fe atoms on the $i$-th nearest-neighbor shell of the considered Mn atom, $Z_{i}$ is the coordination on the $i$-nn shell, and $C_{\rm Fe}$ is the Fe atomic concentration of the system.

\begin{figure}[htbp]
  \centerline{\includegraphics[width=1.0\linewidth]{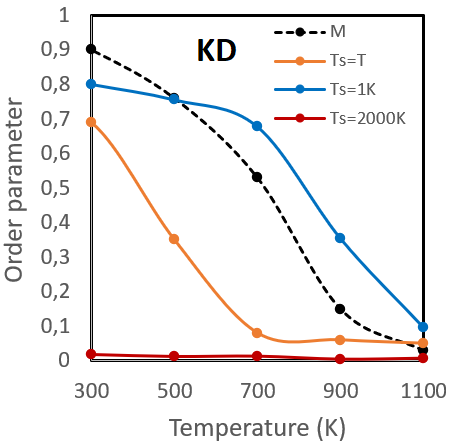}}
  \caption{$1nn$ atomic short range order and reduced magnetization as functions of temperature obtained with the KD-model, at a 10 at. \% Mn concentration.}
  \label{corelsrom}
\end{figure}

The calculated $1nn$-ASRO is shown in Fig. \ref{corelsrom} for a Fe-10 at.\%-Mn alloy. The temperature evolution of the reduced magnetization is also given for information. 
In order to investigate the interplay between the magnetic and chemical orders, similar MC simulations have been performed with a fixed spin temperature $T_s$, independent of the atomic temperature $T$. 
Consistent with Fig.~\ref{0K_mixing}, the low temperature ASRO within the KD model is dominated by the magnetic degrees of freedom. Hence, by imposing a 1K (resp. 2000K) spin temperature, we find a generally larger (resp. lower) ASRO. A similar trend is also observed with ASRO of farther neighboring shells. We therefore confirm that magnetic ordering enhances the unmixing tendency in bcc Fe-Mn alloys.

\smallbreak
\subsection{Vacancy properties near a Mn solute}
\label{vac_properties}

As our KD-model also allows to consider the presence of a vacancy, it can be employed to predict the vacancy-Mn interaction properties, particularly the magnetic free energy of binding (accounting for the magnetic entropy) versus temperature. This value dictates the vacancy concentration around Mn, which is especially important for the determination of solute diffusion coefficients via a vacancy mechanism. \citep{Adda1966, Leclaire1970}

In practice, we calculate the magnetic free energy of formation of a vacancy at a $1nn$ distance of the solute and in a pure Fe lattice, by evaluating an equilibrium vacancy concentration ratio between the system of study at each temperature and a reference system with a known vacancy formation energy (here the perfectly FM bcc Fe) via Monte Carlo simulations, using the same approach as in Ref. \onlinecite{Schneider2020}. A description is given in Sec. \ref{mc_details}. The 1nn Mn-vacancy binding free energy results from the difference between these two formation free energies.

\begin{figure}[htbp]
  \centerline{\includegraphics[width=1.0\linewidth]{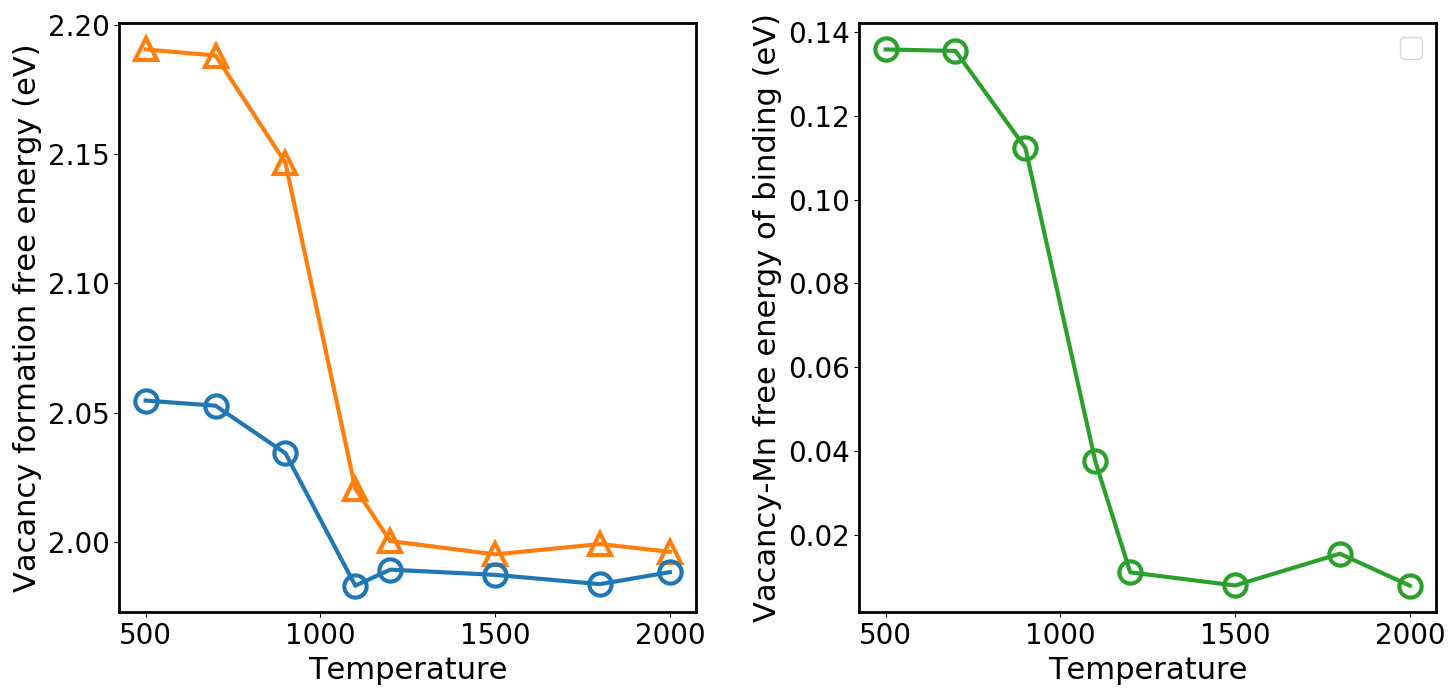}}
  \caption{Left panel: Temperature dependence of the vacancy formation magnetic free energy, in pure Fe (orange) and in the nearest-neighbor shell of a Mn solute (blue). Right panel: Temperature dependence of the vacancy-Mn magnetic free energy of binding. The involved spin-MC simulations adopts a temperature rescaling factor corresponding to a Bose-Einstein statistics, in order to obtain numerical results consistent with the previous study\citep{Schneider2020}. The same re-scaling factor for pure Fe is applied for the extremely dilute Fe-Mn system.}
  \label{form_bind}
\end{figure}

The left panel of Fig. \ref{form_bind} shows the vacancy formation magnetic free energy in pure Fe and at $1nn$ sites of a Mn solute in Fe, as functions of temperature. Concerning the pure Fe case, the vacancy formation magnetic free energy obtained in the low and high temperature regimes (respectively 2.20 and 1.99 eV in FM and PM magnetic states) is in agreement with previous experimental and DFT data from the literature which range from 2.00 to 2.24 eV in the FM state and from 1.54 to 1.98 eV in the PM state \citep{DeSchepper1983,Huang2010, Wen2016, Ding2014, Sandberg2015}. We note that the vacancy formation magnetic free energy in the PM state shows very scattered results in the literature, which are very sensitive to the computational details, while the various studies are very consistent concerning the FM state \citep{DeSchepper1983, Huang2010, Wen2016, Ding2014, Sandberg2015}. As can be noticed, at low temperatures the formation free energy at $1nn$ sites of the solute is approximately 0.14 eV lower than the value in pure Fe, which is consistent with the magnetic free energy of binding obtained via ab-initio calculations. This result is also consistent with a previous first-principles study \citep{Gorbatov2016} which find the Mn-vac binding energy in the FM state to be 0.16 eV. Interestingly, as the temperature increases, this difference decreases towards approximately zero in the fully paramagnetic regime (right panel of Fig. \ref{form_bind}). This solute-vacancy binding behaviour indicates that the magnetic disorder is able to erase the chemical effects in the very dilute alloys. Since we have observed an identical behavior in the case of Cu solutes in Fe\citep{Schneider2020}, while Cu and Mn have very different magnetic properties, it appears to be a general behaviour, independent of the chemical nature of the solute.

\smallbreak
\section{Conclusions}
\label{Conclusions}

Body-centered cubic Fe-Mn alloys present complex and atypical magnetic-interaction behaviours.
Guided by the goal of studying temperature-dependent properties of these alloys, we parameterized an effective interaction model containing explicitly both magnetic and chemical variables. We adopted a knowledge-driven fitting procedure, that is, using only a rather small amount of relevant DFT data. A progressive parameterization strategy was used, which puts emphasis on those data considered to be physically most important. Based on conclusions from DFT studies, one of our assumptions is the dominance of Mn-Mn magnetic interactions over the Fe-Mn ones, in the presence of a magnetic frustration. Therefore, the atypical presence of two magnetic minima for Mn solutes should be correctly captured Further, including non-collinear magnetic configurations in the fitting database turned out to be essential for a satisfactory model.

In order to benchmark the model and its dependence on the knowledge-driven assumptions, it was compared with a second parameterization method. To this end, we used a machine-learning technique based on Ridge regression for the same training data. Apart from the dependence of the starting values, the resulting model considered all the DFT data simultaneously without bias. Though being aware of the insufficient data density for machine learning techniques, this two-fold strategy raised our awareness of strengths and potential imprecision of the knowledge-driven model. It turned out that the magnetic interaction strength is rather independent of the fitting procedure, while the energetic properties such as the
mixing enthalpies are much more sensitive to the model parameters. In fact, the energetic properties in the alloys are highly dependent on the magnetism.

Finite temperature Monte Carlo simulations were then performed in order to show the ability of the KD-model to predict properties which are not included in the fitting data. For instance, the Mn concentration dependence of the magnetic transition temperature is found in excellent agreement with most experimental results. At variance with most experimental methodologies, only providing averaged properties, our approach allows to access the local magnetic moment around individual atoms. It allowed us to
study the temperature evolution of the distribution of the angle between neighboring Fe and Mn spins, along with the evolution of magnitude of Mn magnetic moments. We observed that, contrary to Cr atoms in bcc Fe-Cr alloys, the average magnetic moment of Mn atoms in bcc Fe-Mn does not follow the total magnetization decrease. Indeed, the temperature induced magnetic disordering of Mn atoms is reinforced by the possibility for each spin to switch between the AF-Mn and FM-Mn states.

The temperature dependence of the mixing energy over the whole range of concentration was also determined. The results suggest that the unmixing tendency is highly related to the magnetic order of the system. Moreover, we identified a correlation between the chemical short-range order and the total magnetization of the system. We show that if constraining the spin temperature to asymptotically low or high temperature values highly affects the $1nn$ chemical SRO. This study allowed to further confirm the enhancement of the unmixing tendency by the magnetic ordering.

Finally, we show that it is fully possible to go beyond the ideal defect-free alloys and to consider the presence of a vacancy using such a model. We provided as an example of application the temperature evolution of the vacancy formation magnetic free energy nearby a Mn solute, showing a strong decrease of solute-vacancy binding with the emergence of magnetic disorder. This result is the first key ingredient for the study of Mn solute diffusion in bcc Fe, to which a future paper will be fully dedicated.

\smallbreak
\section{Appendix: Model parameters}
\label{parameters}

The parameters of the two models are given in the following tables.

\begin{table*} [!htbp]
\begin{center}
\begin{tabular}{cccccc}
  \hline
  Distance & $1nn$ & $2nn$ & $3nn$ & $4nn$ & $5nn$ \\
  \hline
  $J_{Fe-Fe}$ & -3.39 & -2.26 & -0.83 & 0.42 & 0.44 \\
    \hline
  $J_{Mn-Mn}$ & 1.51 & 1.30 & 0.26 & -0.98 & 0.53 \\
    \hline
  $J_{0}$ & 0.057 & 0.066 & 0.042 & 0.089 & 0.026 \\
    \hline
  $V_{Fe-Fe}$ & -10.85 & 8.18 & 0 & 0 & 0 \\
    \hline
  $V_{Mn-Mn}$ & 4.63 & -1.93 & -1.06 & -0.19 & 0.25 \\
    \hline
  $V_{Fe-Mn}$ & -6.09 & 3.75 & 0 & 0 & 0 \\
      \hline
  $J_{0}(1nn vac)$ & 0.232 & 0.261 & 0.187 & 0.327 & 0.138 \\
    \hline
  $V_{Fe-Mn}(1nn vac)$ & 30.2 & -17.3 & 0 & 0 & 0 \\
      \hline
  $J_{0}(2nn vac)$ & 0.189 & 0.212 & 0.151 & 0.267 & 0.110 \\
    \hline
  $V_{Fe-Mn}(2nn vac)$ & -30.2 & 17.3 & 0 & 0 & 0 \\
  \hline
\end{tabular}
\caption{\label{femag_interactions}
KD-model: Magnetic and chemical interaction parameters between two atoms, depending on their relative distance, respective species, and the eventual proximity of a vacancy (in meV)}
\end{center}
\end{table*}

\begin{table*} [!htbp]
\begin{center}
\begin{tabular}{ccccccc}
  \hline
  Interactions & $A^{0}$ & $B^{0}$ & $A^{1}$ & $B^{1}$ & $A^{2}$ & $B^{2}$ \\
  \hline
  Fe & -259.0 & 27.6 & -250.2 & 24.7 & -285.7 & 35.2 \\
    \hline
  Mn & -37.70 & 6.93 & -116.2 & 8.54 & -37.70 & 6.93  \\
  \hline
\end{tabular}
\caption{\label{fe_onsite}
KD-model: Magnetic on-site terms of Fe and Mn atoms, depending on the presence or not of a vacancy (in meV)}
\end{center}
\end{table*}

\begin{table*} [!htbp]
\begin{center}
\begin{tabular}{cccccc}
  \hline
  Interaction & $a$ & $b$ & $c$ & $d$ & $e$ \\
  \hline
  Value (meV) & 6.50E-8 & -8.40E-6 & 3.83E-4 & -7.16E-3 & 3.04E-2 \\
  \hline
\end{tabular}
\caption{\label{femn_poly}
KD-model: Parameters of the polynomial function ensuring the local Mn concentration dependence of Fe-Mn magnetic interaction parameters (in meV)}
\end{center}
\end{table*}

\begin{table*} [!htbp]
\begin{center}
\begin{tabular}{cccccc}
  \hline
  Distance & $1nn$ & $2nn$ & $3nn$ & $4nn$ & $5nn$ \\
  \hline
  $J_{Fe-Fe}$ & -14.4 & 0.03 & 2.65 & 0 & 0 \\
    \hline
  $J_{Fe-Mn}$ & -2.23 & 0 & 0 & 0 & 4.14 \\
    \hline
  $J_{Mn-Mn}$ & 24.29 & 16.09 & -8.96 & -10.00 & 10.00 \\
    \hline
  $V_{Fe-Fe}$ & -1927.36 & -0.12 & 0 & 0 & 0 \\
    \hline
  $V_{Mn-Mn}$ & -1188.99 & -1302.10 & -53.32 & -2.86 & 47.48 \\
    \hline
  $V_{Fe-Mn}$ & -1577.47 & -659.95 & 0 & 0 & 0 \\
		\hline
  \hline
\end{tabular}
\caption{\label{femag_interactions}
ML-model: Magnetic and chemical interaction parameters between two atoms, depending on their relative distance, respective species (in meV)}
\end{center}
\end{table*}

\begin{table*} [!htbp]
\begin{center}
\begin{tabular}{cccc}
  \hline
  Interactions & $A$ & $B$ & $C$\\
  \hline
  Fe & -142.70 & 15.64 & 0\\
    \hline
  Mn & -3.83 & -2.22 & 0.34\\
  \hline
\end{tabular}
\caption{\label{fe_onsite}
ML-model: Magnetic on-site terms of Fe and Mn atoms (in meV)}
\end{center}
\end{table*}

\begin{table*} [!htbp]
\begin{center}
\begin{tabular}{ccc}
  \hline
  Interaction & $J_0^{n}$ & $\Delta J_{Fe-Mn}$ \\
  \hline
  Value (meV) & 12.87 & -5.22\\
  \hline
\end{tabular}
\caption{\label{femn_poly}
ML-model: Parameters of the function ensuring the local Mn concentration dependence of Fe-Mn magnetic interaction parameters (in meV)}
\end{center}
\end{table*}

\acknowledgments

The authors would like to thank O. Hedge and Dr. F. Soisson for fruitful discussions. This work was partly supported by the French-German ANR-DFG MAGIKID project (Grant HI1300/13-1). Ab initio calculations were performed using Grand Equipement National de Calcul Intensif (GENCI) resources under the A0070906020 project and the CINECA-MARCONI supercomputer within the SISTEEL project.

\bibliographystyle{apsrev}
\bibliography{Schneider_24july}

\clearpage

\end{document}